\documentclass[aps,a4paper,twocolumn,floatfix,showpacs,
groupaddress]{revtex4}

\usepackage{bm,amsmath,amssymb,amsfonts}

\usepackage[dvips]{graphicx}
\usepackage[colorlinks=true,citecolor=blue,urlcolor=red,filecolor=blue]{hyperref}
\usepackage{color}
\definecolor{mycolor}{rgb}{0.60,0.10,0.40}
\begin{document}

\title{\textcolor{mycolor}
{Analytical study of quasi-one dimensional flat band networks and slow light analogue}}  

\author{Atanu Nandy}
\email{atanunandy1989@gmail.com}
\affiliation{Department of Physics, Kulti College, Kulti, Paschim Bardhaman,
West Bengal-713 343, India}

\begin{abstract}
Exact method of analytical solution of flat, non-dispersive eigenstates in a class of quasi-one dimensional structures is reported within the tight-binding framework. The states are localized over certain sublattice sites. One such finite size cluster of atomic sites is decoupled from the rest of the system by the special `non-permissible' vertex having zero amplitude. This immediately leads to the \textit{self-trapping} of the incoming excitation. We work out an analytical scheme to discern the localizing character of the diffraction free dispersionless modes using real space renormalization group technique. Supportive numerical calculations of spectral 
profile and transport are demonstrated to substantiate the essence of compact localized states.
Possible experimental scope regarding the photonic analogue of the tight-binding electronic case is also discussed elaborately. This eventually unfolds the concepts of \textit{slow light} and the related 
re-entrant mode switching from the study of optical dispersion.
\end{abstract}
\maketitle
\section{Introduction}
\label{intro}

There are several translationally invariant lattices in the tight-binding description that ensure the existence of one (or more) entirely dispersionless flat band(s) in the spectrum and hence are called flat band (FB) networks. Lattice systems containing flat bands have been of great interest over the recent few years. 
In general, the localization phenomenon incidentally happens due to the presence of disorder. This is the celebrated case of Anderson localization~\cite{anderson}. After completion of half century of it, the subject is still alive in many aspects of condensed matter physics. But there are some special low dimensional networks present where excitation can be localized even in absence of any disorder and thus the corresponding single particle eigenstate forms a completely momentum independent flat band in the whole Brillouin zone. 
 Due to the phase cancellation in
the presence of local spatial symmetries wave excitation does not diffuse beyond the 
periphery of a finite sublattice sites in case of several low dimensional networks.
This brings the essence of \textit{self-localization}.
This is in complete contrast to the canonical case of Anderson localization where exponential
localization is caused by disordered environment.
Comparatively, \textit{compact localized states} (CLS) typically occur in overall periodic systems.
Such CLS are used for the formation of tunable, local symmetry-induced bound states in an energy
continuum.
The key fact is the destructive kind of quantum interference resulting from the local topological symmetry present in the underlying geometry. This also implies the
 well-known flat band localization where the quenched kinetic energy of the particle 
 leads to complete immobility of the electron. 
The specific topology of the system has a strong influence on the overall spectrum and in most of the cases, it induces remarkable spectral features introducing dispersionless flat bands in several low-dimensional networks.

The journey of flat band network started almost thirty years ago from Sutherland's prediction~\cite{suther}
of flat band and the corresponding state is ``strictly localized" in the dice lattice (which relates to previous research on quasi-periodic lattices and Penrose tilings).
This section of study had an initial interest in the field of strongly correlated systems because of having and open platform in the context of ferromagnetism~\cite{lieb}-\cite{tasaki2}. The nearly FB states having non-zero Chern number supports an interesting physics regarding the fractional quantum hall effect~\cite{kapit}-\cite{chamon}. 
Starting from all these things this area has extended its impact on several branches of condensed matter physics. Extensive studies have been done related to gapped FB states or gapless chiral modes in graphenes~\cite{yao}, optical lattices of ultracold atoms~\cite{bloch}, wave guide arrays~\cite{lede} or in microcavities having exciton-polaritons~\cite{kim}.

Several tight binding lattices are studied in respect of this flat band localization. Kagom\'{e}, Lieb, Diamond, $AB_2$ stub, and sawtooth lattices are some vivid examples of the flat band lattices, and some 
general methods~\cite{dias}-\cite{flach3} are proposed to design more lattice structures with flat bands. Many theoretical works related to the flat band lattices have been done in the context of ferromagnetism~\cite{shen}, superconductivity~\cite{sarma}
 and Wigner crystals~\cite{kawa}-\cite{kopnin}. Theoretical understanding is based on simplified models or approximations and comparison with experiment is crucial.

Recent interest in this area has arisen due to experimental realization of flat band systems~\cite{seba,vicen}
 in Lieb photonic lattices and tight-binding photonic bands in metallophotonic wave guide networks~\cite{oka}.
  Such flat photonic bands are intimately connected to the path-breaking idea of engineering slow light~\cite{atanu4}
   with low group velocity which opens up the possibility of ``spatial compression of light energy". 
The most intriguing property of the flat band is the localization of the wave function at certain sub-lattice sites. The single particle wave function is finite within a cluster of atomic sites but becomes zero at some connecting nodes, making it impossible to escape from those finite size clusters of atomic sites. The quenched kinetic energy suppresses the quantum transport of the wave train; the group velocity vanishes. This immobility corresponding to such \textit{self-localized} modes eventually contribute to non-dispersive part of the energy-wave vector relation. The singular behavior arising out of the immobility is expected to produce anomalous behavior in the physical properties, transport and optical responses.
Another interesting fact about FB is that the dispersion curve is essentially flat in curvature for that specific energy. This implies that the single particle energy state is independent of the momentum of the particle. This brings a concept of divergent effective mass tensor i.e., the particle behaves like a super-heavy one and the immediate immobility severely affects the transport of quantum mechanical wave packet that results in some unconventional phenomena such as inverse Anderson transition~\cite{goda}.
Moreover these FB systems can be classified into two separate categories according to the response to the external perturbation, particularly uniform magnetic flux or any kind of disorder effect. 
This first category~\cite{miel4}-\cite{atanu2} do not display flat bands for finite magnetic flux. 
In contrast, Lieb class of lattices exhibit robustness of the FB i.e., the FB remains unperturbed in respect of application of external magnetic flux.

In a flat band, diffraction is totally suppressed due to destructive interference, in a way analogous to geometric frustration, giving rise to eigenmodes that are compactly localized in space. Contrary to the canonical case of Anderson localization, for those compact localized states, wave excitations strictly vanish outside a finite size sublattice of a system and this is entirely caused by destructive interference in the presence of local spatial symmetries. The immediate application of CLS lies in the information transmission~\cite{mejia}-
\cite{xia} and directly stems from their compactness.
Beacuse of extremely low mobility of the wave packet CLS does not spread out spatially during evolution.
CLSs are suitable candidates for the transmission of signal along photonic wave guide arrays avoiding `crosstalk' between wave guides~\cite{mejia2}.
Moreover, CLSs essentially enable the appearance of isolated bound states within 
a scattering continuum~\cite{von,herr}.

The principal motivation behind this article is to look into possibility of demonstration of a general analytical scheme within the tight-binding formalism to discern the flat non-dispersive states for quasi-one dimensional structures. The states are essentially localized over clusters of atomic sites and one such trapping cluster is effectively isolated from the rest of the network. This is partially by the destructive quantum interference and partially by the physical boundary formed by the sites with vanishing amplitude. We have discussed a decoupling technique to analyze those self-localized pinned states where there is no overlap between the wave functions of the neighboring nodes corresponding to such energy. This confirms the localized nature of the dispersionless modes.
 
The tight-binding analogy with the electronic case and the corresponding photonic scenario helps us to propose a monomode wave guide network made of same geometry. We present here a simple analytical method to detect the sharply localized photonic eigenmodes that are pinned on certain atoms or atomic clusters in a periodic array of diamonds. The non-dispersive character of such states is explicitly worked out. The other point of interest in the present work is to analyze the distribution of the amplitudes of such flat-band states in real space.

We find interesting results. We have been able to work out a method to extract the flat band modes for a set of quasi-one dimensional structures. The non-diffusive nature of such non-dispersive states is also confirmed. The presence of both dispersionless and dispersive states leads to an interesting situation of mode crossover.

Before we end this section, it is worth mentioning that single-mode wave guides have recently been fabricated and used in a quasiperiodic optical setup~\cite{ringel} to unravel topological states in quasicrystals. Localization transition in one-dimensional quasiperiodic lattices has also been investigated experimentally in 
recent times~\cite{sorel}. Also quite recently, the computational design of flat band models~\cite{ochi,hase}
 has enriched
this section of study. The present proposal, to our mind, can thus be tested with an appropriately designed wave guide network.
\section{Description of model and analysis: Rhombic lattice}
\label{cop}

\subsection{Spectral Information by RSRG method}
\label{density}
We start our discussion with the prototype example of $AB_2$ rhombic lattice. 
The electron's hopping is restricted between the nearest neighboring sites only. According to the number of nearest neighbors we distinguish between
\begin{figure}[ht]
\centering
\includegraphics[clip,width=6cm,angle=0]{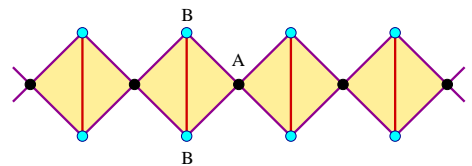}
\caption{(Color online) A portion of an infinitely long quasi-one dimensional $AB_2$ rhombic lattice.}  
\label{lattice1}
\end{figure}
 two types of sites, viz., one with coordination number three and colored in blue (named as $B$ sites) and the other, black circles, having a coordination number four ($A$ sites). Spinless non-interacting electrons are described by the tight-binding Hamiltonian in Wannier basis,
\begin{equation}
H  = \sum_{m} \epsilon_m c_{m}^{\dagger} c_{m}
+\sum_{\langle mn \rangle} t_{mn} \left[c_{m}^{\dagger} 
c_{n} 
+ h.c. \right] 
\label{hamiltonian}
\end{equation}
In the above expression $\epsilon_m$ is the on-site energy term at the respective quantum dot location that gives the potential contribution and $t_{mn}$ is the nearest neighbor hopping parameter (or \textit{overlap integral}) carrying the kinetic signature. Depending on the coordination number $\epsilon_m$ can assume values equal to
 $\epsilon_3$ or $\epsilon_4$ according to the local connection. It is needless to say that without any loss of generality we just assign the numerical values of $\epsilon_3$ and $\epsilon_4$ both equal to zero throughout the analytical calculation since we are interested to see the effect of topology of the lattice. The vertical connection (shown as solid line) between the top and the down vertices is taken as $\lambda$. The difference equation which is an alternative discretized form of the Schr\"{o}dinger's equation can be written as follows
\begin{equation}
( E - \epsilon_m ) \psi_m = \sum_{n} t_{mn} \psi_{n}
\label{difference}
\end{equation}

\subsection{Analytical construction of compact localized state}
\label{compact}
\begin{figure}[ht]
\centering
\includegraphics[clip,width=6cm,angle=0]{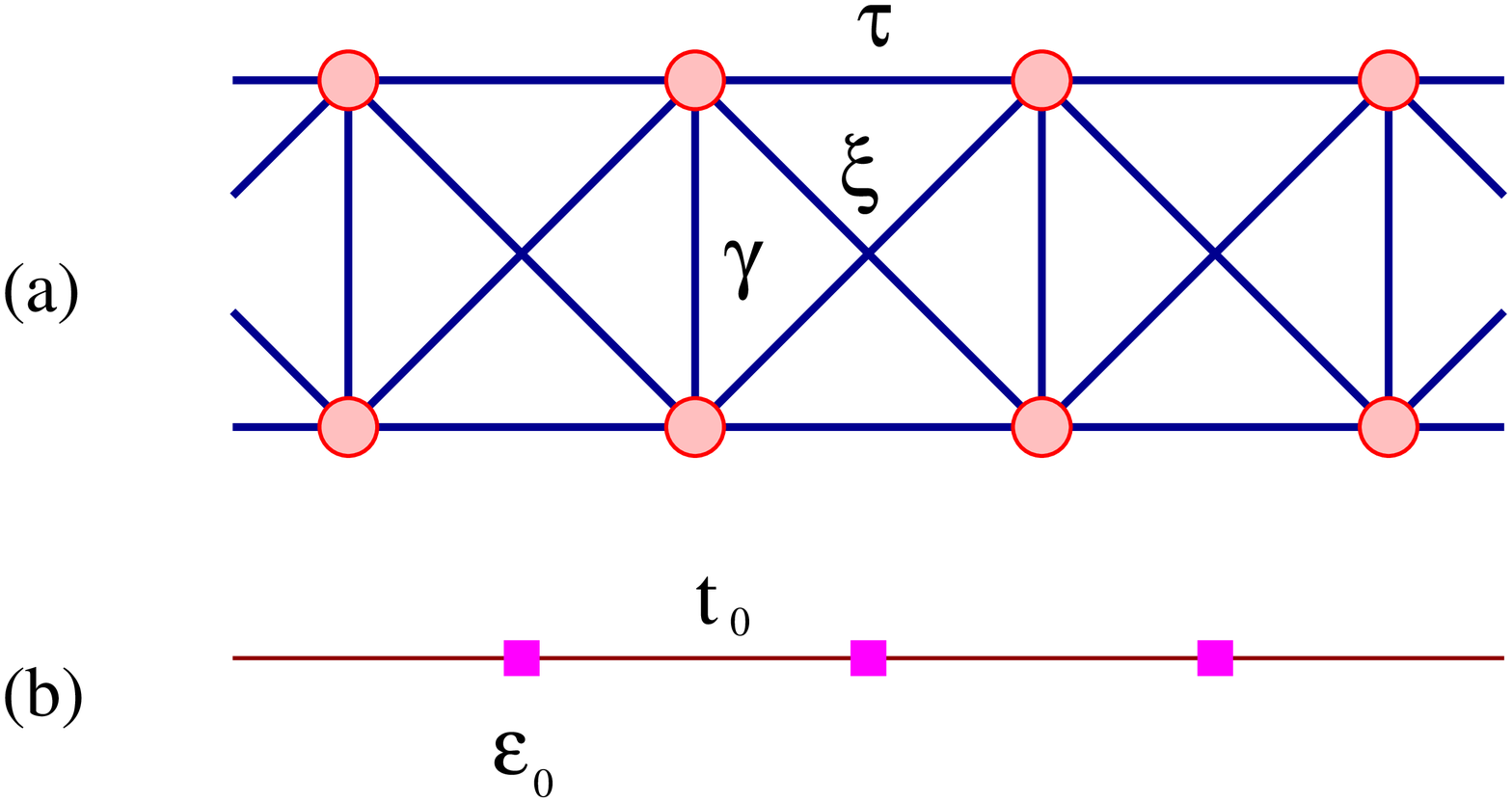}
\caption{(Color online) (a) The \textit{effective} two-arm ladder with the \textit{renormalized} parameters
(all the sites have the same on-site potentials $\tilde{\epsilon}$) and 
(b) The \textit{renormalized} periodic linear chain of identical atomic sites. This scheme helps to compute
the dispersion relation.}  
\label{ladder1}
\end{figure}

We will follow the real space renormalization group technique to discern the non-dispersive \textit{compact localized} state. By virtue of the above expression one can easily decimate out the central vertices in terms of the amplitudes of the surviving nodes to map \textit{effectively} it into a two-arm 
ladder network (Fig.~\ref{ladder1}(a))
 comprising
 identical atomic nodes with `effective' on-site potential $\tilde{\epsilon}=\epsilon_3+\frac{2 t^2}{(E-\epsilon_4)}$. The overlap integrals will also be renormalized and function of energy. The hopping along each arm of the ladder now becomes $\tau= t^2/(E-\epsilon_4) $ and the inter-arm vertical connection becomes $\gamma= \lambda+2t^2/(E-\epsilon_4)$. This decimation procedure following RSRG technique creates a second neighbor
  connection i.e., diagonal hopping integral $\chi= t^2/(E-\epsilon_4)$.

With all these \textit{renormalized} parameters we can now rewrite the difference equation for this effective two-legged ladder geometry. The equation can be written as,
\begin{widetext}
\begin{eqnarray}
(E - \tilde{\epsilon}) \psi_{n,A} & = & \tau (\psi_{n+1,A} + \psi_{n-1,A}) + 
\gamma \psi_{n,B} + \xi (\psi_{n+1,B} + \psi_{n-1,B}) \nonumber \\
(E - \tilde{\epsilon}) \psi_{n,B} & = & \tau (\psi_{n+1,B} + \psi_{n-1,B}) + 
\gamma \psi_{n,A} + \xi (\psi_{n+1,A} + \psi_{n-1,A}) 
\label{difflad}
\end{eqnarray}
\end{widetext}
The above equation for the coupled system can be easily cast in a composite matrix form using the potential and hopping matrices in the form,
\begin{widetext}
\begin{eqnarray}
\left [
\left( \begin{array}{cccc}
E & 0 \\
0 & E
\end{array}
\right ) - 
\left( \begin{array}{cccc}
\tilde\epsilon & \gamma \\
\gamma & \tilde\epsilon
\end{array}
\right)
\right ]
\left ( \begin{array}{c}
\psi_{n,A} \\
\psi_{n,B}  
\end{array} \right )
& = & 
\left( \begin{array}{cccc}
\tau & \xi \\ 
\xi & \tau 
\end{array} 
\right)
\left ( \begin{array}{c}
\psi_{n+1,A} \\
\psi_{n+1,B}  
\end{array} \right )
+
\left( \begin{array}{cccc}
\tau & \xi \\                                                  
\xi & \tau
\end{array}
\right)
\left ( \begin{array}{c}
\psi_{n-1,A} \\
\psi_{n-1,B}
\end{array} \right )
\label{eqladder}
\end{eqnarray}
\end{widetext}
Now here one important point to mention that the forms of the potential matrix (comprising the on-site energy and vertical connection) and the hopping matrix (connecting $\tau$ and $\chi$) are such that they will commute, and hence can be simultaneously diagonalized by a similarity transform.
So taking the advantage of this commutation, we can make a uniform change of basis defined by $\phi_n = \bm{M}^{-1} \psi_n$ and the coupled difference equation for ladder can now be easily decoupled into a set of two independent difference equations for the two arms. The matrix $M$ makes both the potential and hopping matrices diagonal. The decoupled set of equations are now obviously free from any coupling term (that makes the inter-arm connection) and reads in terms of the original lattice parameters as,
\begin{widetext}
\begin{eqnarray}
\left [ E - \left (\epsilon_3 + \lambda + \frac{4t^2}{E-\epsilon_4} \right ) \right ] 
\phi_{n,1} & = & \frac{2t^2}{E-\epsilon_4} ( \phi_{n+1,1} + \phi_{n-1,1} ) \nonumber \\
\left[ E - \left(\epsilon_3 - \lambda \right) \right] \phi_{n,2} & = & 0
\label{decouplep}
\end{eqnarray}
\end{widetext}
The first one immediately represents a perfectly ordered chain of identical atomic sites with energy dependent renormalized on-site potential $\epsilon_3 + \lambda + \frac{4 t^2}{(E-\epsilon_4)}$ and the effective overlap parameter $t = \frac{4 t^2}{(E-\epsilon_4)}$. The second equation represents the difference equation (in the new basis) of an \textit{isolated atom} with effective on-site energy. The corresponding wave function does not have any overlap with the wave function envelope of the neighboring nodes, i.e., completely isolated from the rest of the system. The eigenstate is a \textit{compact localized state} (CLS), 
in the spirit of S. Flach et. al.~\cite{flach4},
which is localized partially by the destructive kind of quantum interference and partially by the physical boundary formed by the sites with zero amplitudes. The topology of the system concerned is the key reason for happening the perfect geometric phase cancellation. This leads to an eigenfunction with amplitudes \textit{pinned} at the top and down vertices and the eigenvalue corresponding to this pinned localized state is
\begin{figure}[ht]
\centering
\includegraphics[clip,width=6cm,angle=0]{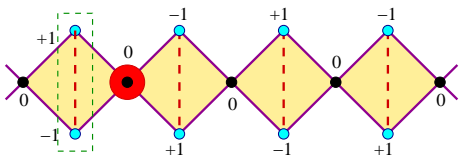}
\caption{(Color online) Amplitude distribution for the \textit{flat band} state at $ E = \epsilon_3 - \lambda$
. The green encircled area shows the characteristic \textit{trapping cell} and the red circle shows the
zone that is not permissible for electron because of the nodes having zero amplitude.}  
\label{amp1}
\end{figure}
 $E = \epsilon_3 - \lambda$. The amplitude distribution is shown pictorially in the Fig.~\ref{amp1}. The incoming wave train having this particular energy will be trapped in the local clusters (vertices with coordination number equal to three) and will show no evolution dynamics. This macroscopically degenerate states gives a momentum independent contribution in the $E-k$ dispersion curve and is therefore called a flat band (FB) state~\cite{mati}, 
 a locus of zero mobility points in the momentum space near the Fermi level.
\subsection{Density of states} 
\label{spec}
Before looking for the non-dispersive nature of the 
compact localized state supported by this quasi-one dimensional structure, it is generally advisable to have the overall idea of the allowed eigenspectrum for such a system. The first one of the above equation (Eq.~\eqref{decouplep})
 represents a perfectly ordered chain of identical atomic sites with energy-dependent on-site potential and hopping parameter. The second one depicts an ``\textit{atomic-like}" state. Each equation has its own density of states spectrum and obviously a convolution of these two will bring the true density of eigenstates of the entire system.

\begin{figure}[ht]
\centering
\includegraphics[clip,width=6 cm,angle=0]{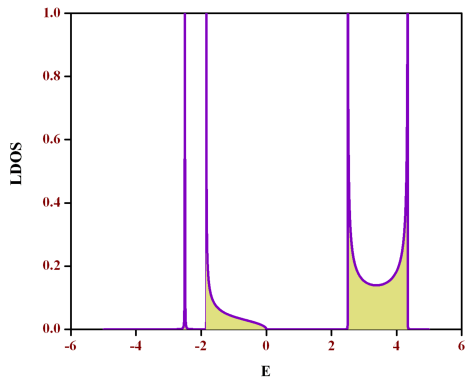}
\caption{(Color online) Local electronic density of states profile as a function of energy. We have set
the numerical values of all the on-site potentials equal to zero, 
i.e., $\epsilon_3 = \epsilon_4 = 0$} and $t = 1$, $\lambda = 2.5$. 
\label{dos1}
\end{figure}

We have however followed an RSRG decimation scheme on the renmormalized ladder (Fig.~\ref{ladder1}(a)).
The renormalization scheme is governed by the following recursion relation
\begin{eqnarray}
\tilde{\epsilon}(n+1) & = & \tilde{\epsilon}(n) + \frac{2}{\delta(n)} (f_1 + f_2) \nonumber \\
\tau(n+1) & = & \frac{1}{\delta(n)} (f_1 + f_2) \nonumber \\
\chi(n+1) & = & \frac{1}{\delta(n)} (f_3 + f_4) \nonumber \\
\gamma(n+1) & = & \gamma(n) + \frac{1}{\delta(n)} (f_3 + f_4)
\label{recursion}
\end{eqnarray}
where $f_1 = [(E-\tilde{\epsilon}(n)) (\tau^{2}(n) + \chi^{2}(n))]$, $f_2 = 2 \tau(n) \chi(n) \gamma(n)$
$f_3 = 2 (E-\tilde{\epsilon}(n)) \tau(n) \chi(n)$,  $f_4 = \gamma(n) (\tau^{2}(n) + \chi^{2}(n))$
and $\delta(n) = (E-\tilde{\epsilon}(n))^2 - \gamma^2(n)$.
After successive steps of renormalization the potential reaches its fixed point value. 
 From this one can calculate the local green's function for the up and down vertices. 
The basic definition of Green's function~\cite{greenfunc} is as follows,
\begin{equation}
G = \sum_{k} \left( E - H \right) ^{-1} | k \rangle \langle k |
\label{green}
\end{equation}
From this expression one can easily compute the local Green's function as
\begin{equation}
G_{00} (E + i \eta) = \frac{1}{N} \sum_{k} \frac{1}{\left( E +i \eta - \epsilon^{\ast} (k) \right)}
\label{green2}
\end{equation}
where $\ast$ mark denotes the fixed point value of the parameter and $\eta$ is
a small quantity added for the numerical calculation of DOS.
 From that green's function calculation we just evaluate the local density of states using the following standard expression
\begin{equation}
\rho (E) =- \left( \frac{1}{\pi} \right) \lim_{\eta\rightarrow 0} Im G_{00} (E + i \eta)
\label{dos}
\end{equation}
The variation of electronic density of states against energy of the electron is plotted in the Fig.~\ref{dos1}. It consists of two separate absolutely continuous sbubands along with the coexistence of an \textit{isolated}
 flat localized mode situated outside the bands. The nature of eigenstates can be easily confirmed if we just look out for the flow of overlap integral under successive RSRG iterations. A careful check of this flow pattern shows that for any energy belonging to the those absolutely continuous regimes the hopping integral never goes to zero, rather follows an oscillating behavior for almost an indefinite number of iteration loops. This is a clear indication that all the states residing inside the continua are all extended in nature. While for the pinned localized FB mode situated at the flank of the spectrum, the overlap parameter immediately converges to zero after few RSRG steps. This means that the overlap of the wavefunction with the nearest neighboring sites decays gradually. This brings a floavor of localization corresponding to $E = \epsilon_3 - \lambda$. 

\subsection{Dispersion relation}
\label{dis1}
 
\begin{figure}[ht]
\centering
\includegraphics[clip,width=6 cm,angle=0]{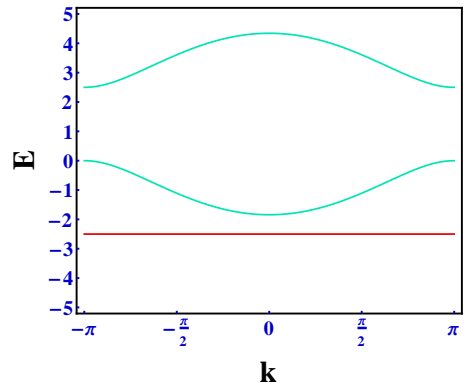}
\caption{(Color online) The energy ($E$) vs. momentum ($k$) relation for the rhombic lattice.
The non-dispersive signature of the FB state is apparent here.}  
\label{pyrodisp}
\end{figure}

Within the tight-binding framework following the RSRG method we have made it possible to confirm the non-dispersive character of that CLS. We refer to Fig.1(c). The top and the down vertices with coordination number equal to three can thus be eliminated to form an effective linear chain of identical atomic sites. Each of the atomic sites can have the on-site potential and the nearest neighbor hopping parameter of the form
\begin{eqnarray}
\epsilon_0 & = & \epsilon_4 + \frac{4 t^2 (E - \epsilon_3 + \lambda)}
{( E - \epsilon_3 )^2 - \lambda^2 } \nonumber \\
t_0 & = & \frac{2 t^2 ( E - \epsilon_3 + \lambda)}{( E - \epsilon_3 )^2 - 
\lambda^2 }
\label{linear}
\end{eqnarray}
With these effective parameters we can now easily write the tight-binding dispersion relation $E = \epsilon_0 + 2 t_0 \cos ka$ which on simplification gives,
\begin{equation}
( E -\epsilon_3 + \lambda) \left [ (E-\epsilon_4) (E-\epsilon_3-\lambda) 
- 8 t^{2} \cos^2 \left(\frac{ka}{2} \right) \right ] = 0
\label{disp1}
\end{equation}
From the above equation we can see the momentum independent part that gives a dispersionless FB at $E = \epsilon_3 - \lambda$. This is in accordance with the result obtained by analyzing the decoupled set of equations. The dispersion relation is plotted in the Fig.~\ref{pyrodisp}.

From the second decoupled equation we see that there is no overlap connection 
\begin{table}[ht]
\caption{Divergence of density of states for the FB mode}
\centering
\begin{tabular}{c c} 
\hline\hline 
Imaginary part of energy ($\eta$) & LDOS ($\rho$) \\ [0.5ex]
\hline
$10^{-2}$ & $15.92$ \\
$10^{-3}$ & $159.15$ \\
$10^{-4}$ & $1591.55$ \\
$10^{-5}$ & $15915.49$ \\
$10^{-6}$ & $159157.99$ \\
$10^{-7}$ & $1588266.91$ \\ [1ex]
\hline
\end{tabular}
\label{table:diverge}
\end{table}
with the nearest neighboring sites the incoming particle having such FB energy
will lose its mobility. 
As the kinetic 
energy of the associated wave packet is quenched, the 
density of states diverges due the relationship
\begin{equation}
\rho = \int v_{g}^{-1} d^3 k
\end{equation}
Such divergences in the spectral profile
are expected to produce anomalous behaviors in 
physical properties as well as transport phenomena and optical 
response. The singularity of spectral profile
with the gradual decrease of imaginary part added to the energy for
the rhombic structure is shown in the adjacent tabular form. 

\subsection{Transmission characteristics} 
\label{tr1}
For the sake of completeness of the above discussion we have also computed the two-terminal transport characteristics of the network under study. The formalism is quite standard and is often used to evaluate the
\begin{figure}[ht]
\centering
\includegraphics[clip,width=6 cm,angle=0]{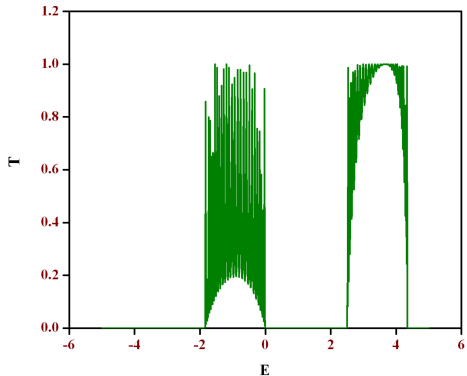}
\caption{(Color online) Two-terminal transport characteristics against energy of the electron.
We have taken
the numerical values of all the on-site potentials equal to zero,
i.e., $\epsilon_3 = \epsilon_4 = 0$} and $t = 1$, $\lambda = 2.5$. 
\label{trans1}
\end{figure}
 transport profile of several quasi-one dimensional networks. The fundamental concept is that we have to clamp our system of finite size in between a pair of semi-infinite periodic leads, the so called `source' and the `drain' with the corresponding lead parameters $\epsilon_0$ and $t_0$. The finite sized network sandwiched in between these two ordered leads is then successively renormalized to reduce it to an effective diatomic molecule (dimer). The two renormalized atomic sites of the dimer molecule will then have energy dependent on-site potentials and overlap parameters containing the information of the system concerned.
The transmission coefficient of the 
lead-network-lead system then is given by a well-known formula~\cite{liu},
\begin{eqnarray}
& T=\frac{4\sin^{2}ka}{|\mathcal{A}|^{2}+|\mathcal{B}|^{2}} \\
&\textrm{with,}\quad \mathcal{A}=[(M_{12}-M_{21})+(M_{11}-M_{22})\cos ka] \nonumber\\
&\textrm{and}\quad \mathcal{B}=[(M_{11}+M_{22})\sin ka]\nonumber
\end{eqnarray}

where, $M_{ij}$ refer to the dimer-matrix elements, written
appropriately in terms of the on-site potentials of the final
renormalized left (L) and right (R) atoms at the extremities of
the finite sized network and the renormalized hopping between
them. $\cos ka = \frac{(E-\epsilon_0)}{2 t_0}$ and $a$ is the lattice constant in
the leads which is set equal to unity throughout the calculation.

For the selective regimes of energy the transparent character of the
system is justified by the high transmitting behavior corresponding to
those regions. The extended nature of the eigenstates and the consequent chaotic
oscillation of the flow of hopping integral is the reason of getting high conduction.
Another fact is that the transmission shows considerably low value corresponding
to the flat band energy as stated above which is expected.

\section{Description of model and analysis: Dimer-plaquette chain}
\label{dimer}
\begin{figure}[ht]
\centering
\includegraphics[clip,width=6 cm,angle=0]{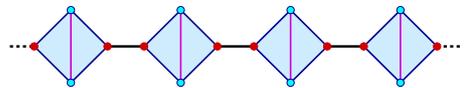}
\caption{(Color online) An infinite array of quasi-one dimensional dimer-plaquette chain.
All the sites have the same on-site parameter $\epsilon$ and inter-arm hopping is $t$ and the 
vertical connection is set as $\lambda$.}  
\label{lattice2}
\end{figure}

The second example of such network we have taken is the dimer-plaquette chain as shown in the Fig.~\ref{lattice2}. The network is well studied in respect of Frustrated quantum Heisenberg antiferromagnetic
 system~\cite{derz} under the influence of a high magnetic field. With a suitable Hamiltonian describing the above network they have shown the existence of dispersionless localized-magnon band. The state is localized within a characteristic \textit{trapping cell} by virtue of phase cancellation. Inspired by this pioneering work we have taken this geometry to work out the same analytical scheme to discern the flat electronic band state following the real space renormalization group method.

In this network there exists only one type of atomic site having number of nearest neighbor equal to three. 
So without any loss of generality, we take the on-site potential of all those sites as $\epsilon$. The hopping integral along the arm of the diamond plaquette and the along the dimer are taken as uniform and assigned as $t$, but the connection between the top and down vertices of each elementary diamond plaquette is considered as different than the others and is taken as $\lambda$.

\subsection{Analytical construction of CLS}

Following the same technique one can easily caste the above network into an \textit{effective} two-arm ladder
geometry with the corresponding renormalized parameters. In this case to get the effective ladder, one has to \textit{eliminate} the two atomic sites of the dimers of the network. By virtue of this decimation process,
the potential matrix now becomes,
\begin{equation}
\bm{\epsilon}=
\left( \begin{array}{cccc}
\epsilon + \frac{2 t^{2} \left( E- \epsilon \right)}{\Delta} & \lambda + \frac{2 t^{2} \left( E- \epsilon \right)}{\Delta} \\
\lambda + \frac{2 t^{2} \left( E- \epsilon \right)}{\Delta} & \epsilon + \frac{2 t^{2} \left( E- \epsilon \right)}{\Delta} 
\end{array}
\right)
\label{mat1}
\end{equation}
Also the hopping matrix reads as,
\begin{equation}
\bm{t}=
\frac{t^3}{\Delta}
\left( \begin{array}{cccc}
1 & 1 \\
1 & 1
\end{array}
\right)
\label{mat2}
\end{equation}
where, $\Delta = \left(E - \epsilon \right)^2 - t^2$. 
Here also, the potential matrix and the hopping matrix commute with each other. So following the same
method, one can now decouple the composite form of the difference equation for the ladder.
We will then get two linearly independent equations (in the changed basis) which are as follows,
\begin{widetext}
\begin{eqnarray}
\left [ E - \left (\epsilon + \lambda + \frac{4t^2 (E-\epsilon)}{\Delta} \right ) \right ] 
\phi_{n,1} & = & \frac{4t^3}{\Delta} ( \phi_{n+1,1} + \phi_{n-1,1} ) \nonumber \\
\left[ E - \left(\epsilon_3 - \lambda \right) \right] \phi_{n,2} & = & 0
\label{decouple}
\end{eqnarray}
\end{widetext}

\subsection{Density of states and transport}
The first one corresponds to an effective linear chain of identical atomic sites with 
the renormalized on-site potential and hopping integral. The spectrum for this chain will extend
from $\epsilon + \lambda + \frac{4t^2 (E-\epsilon)}{\Delta} - \frac{8t^3}{\Delta}$ to 
$\epsilon + \lambda + \frac{4t^2 (E-\epsilon)}{\Delta} + \frac{8t^3}{\Delta}$ and the bands will
be populated by extended kind of eigenfunctions. This resonant character of the bands can be 
easily confirmed if we just check the flow of overlap parameter for any energy belonging to
the \textit{absolutely continuous} regime. It is seen that the flow of hopping 
never converges to zero for an indefinite number RSRG steps which is clear indicative signature of
the states being extended. The other equation again represents an \textit{atomic-like} 
pinned localized state for which the characteristic trapping cell is the top-down atomic dimer.
\begin{figure}[ht]
\centering
(a)\includegraphics[clip,width=6cm,angle=0]{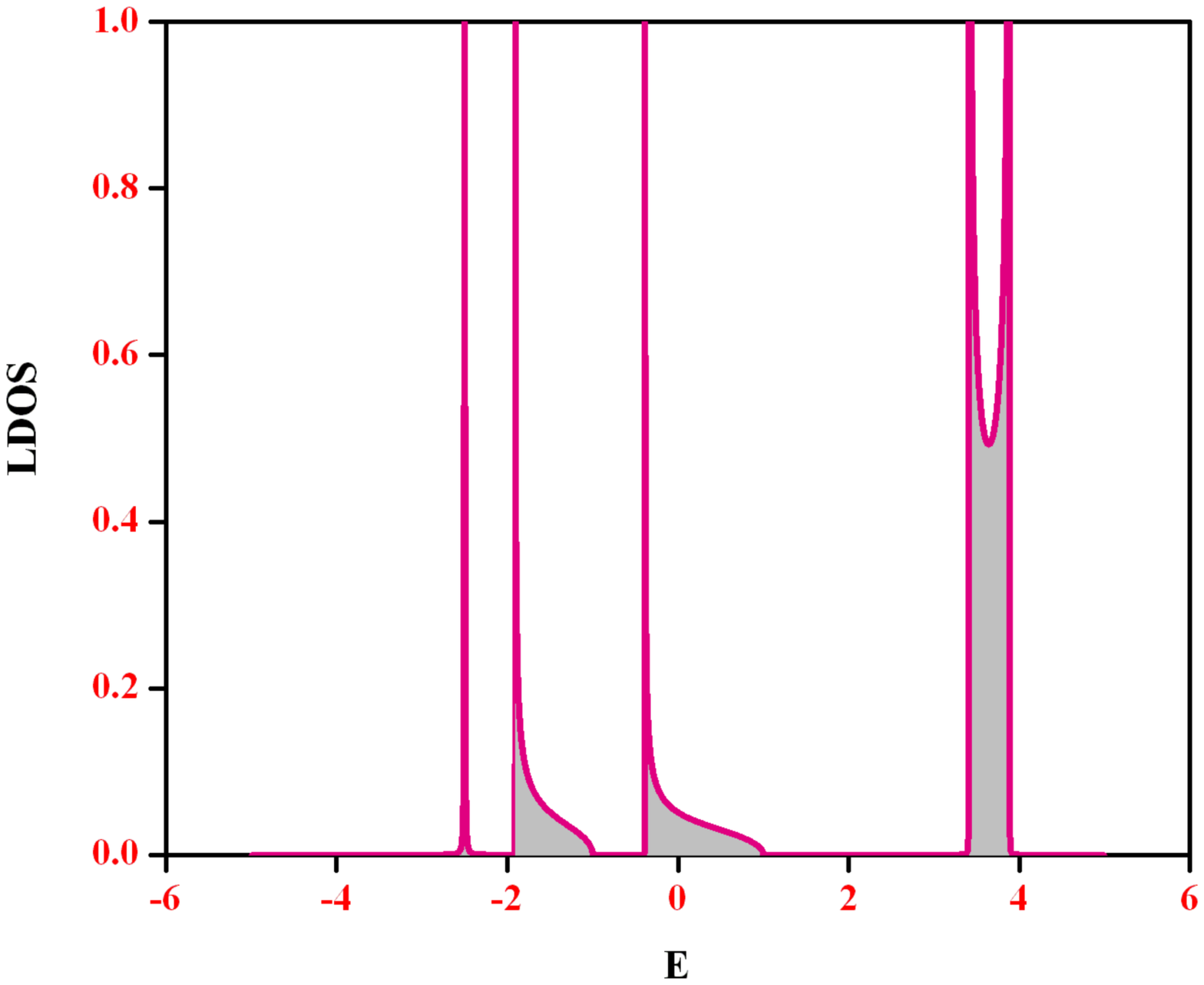}
(b)\includegraphics[clip,width=6cm,angle=0]{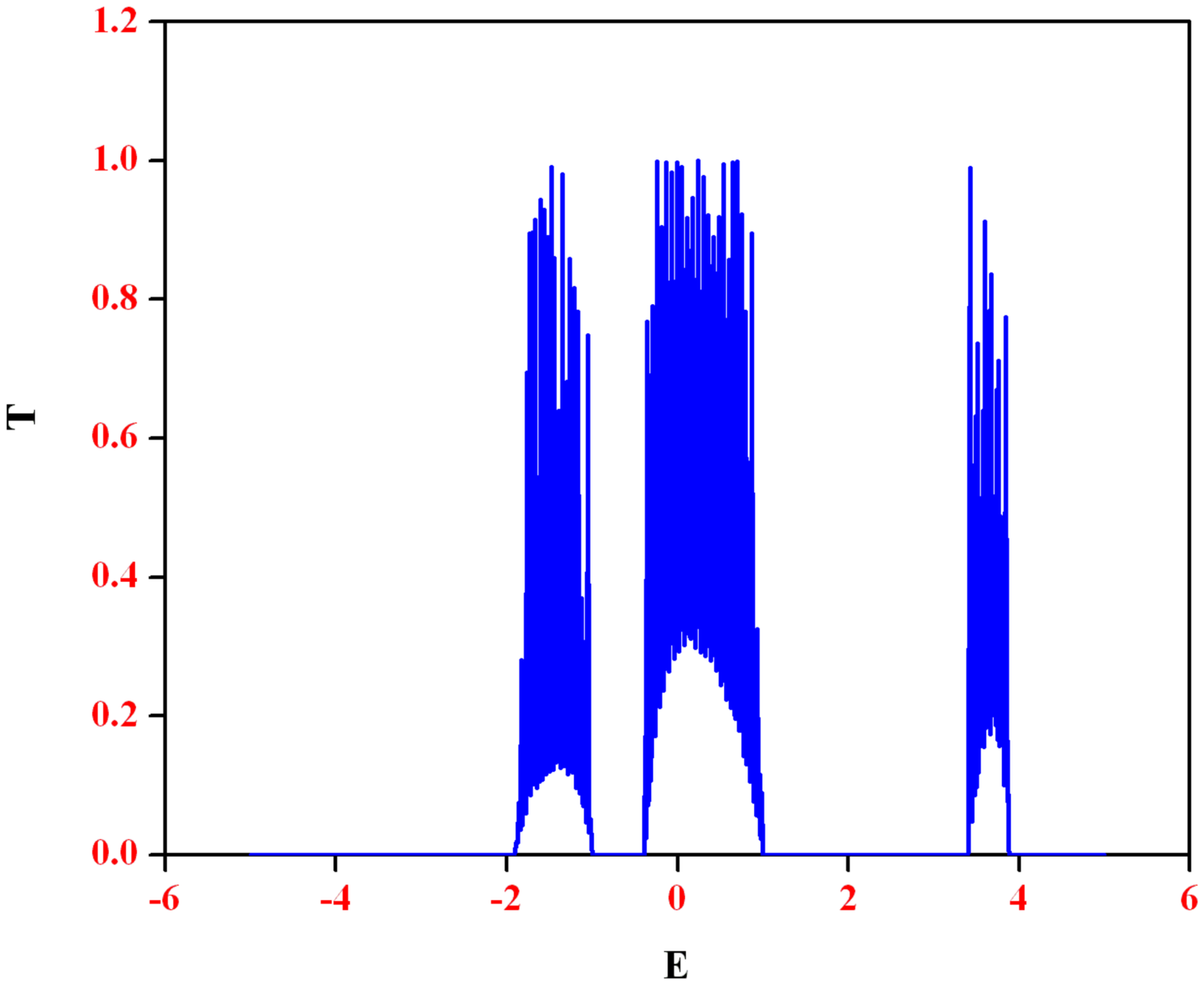}
\caption{(Color online) 
(a) Local spectral landscape of dimer-plaquette chain as a function of energy. We have set
the numerical values of all the on-site potentials equal to zero,
i.e., $\epsilon_3 = \epsilon_4 = 0$} and $t = 1$, $\lambda = 2.5$ and
(b) transmission profile against energy for the same network under the same parametric condition.
\label{dimerspec}
\end{figure}
The amplitudes are confined in the top and down vertices. Because of the zero amplitudes
at the red colored
vertices the particle having such specific FB 
energy, will be confined within the finite size cluster. This brings the flavor of 
the same \textit{compact localized state}.
\begin{figure}[ht]
\centering
\includegraphics[clip,width=6 cm,angle=0]{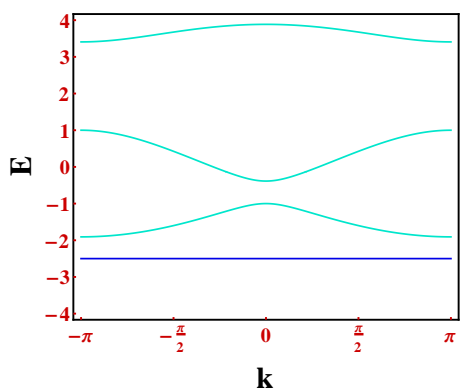}
\caption{(Color online) Dispersion relation of dimer-plaquette chain.}  
\label{dimerdisp}
\end{figure}

\subsection{Dispersion relation}
To justify the non-dispersive character of this FB state constructed analytically, we
will convert the structure into an effective linear chain. For this we just have to eliminate out the
top and down vertices in terms of the amplitudes of the surviving sites.
The tight-binding chain constructed in the above process has the renormalized parameters as follows,
\begin{eqnarray}
\epsilon_0 & = & \tilde{\epsilon} + \frac{\left( \tilde{t}^2 + t^2 \right)}{(E - \tilde{\epsilon})} \nonumber \\
t_0 & = & \frac{t \tilde{t}}{(E - \tilde{\epsilon})}
\label{linear2}
\end{eqnarray}
where, $\tilde{\epsilon} = \epsilon + \frac{2 \left[ \left( E - \epsilon \right) t^2 + t^2 \lambda\right]}{(E - \epsilon)^2 - \lambda^2}$ and $\tilde{t} = \frac{2 \left[ \left( E - \epsilon \right) t^2 + t^2 \lambda\right]}{(E - \epsilon)^2 - \lambda^2}$. 
Using the above parameters one can have the form of the dispersion relation in terms of the parameters
of the original lattice as given by,
\begin{widetext}
\begin{equation}
( E -\epsilon + \lambda) \left[ (E-\epsilon) \left( (E- \epsilon) (E- \epsilon - \lambda) -4 t^2 \right) 
- t^2 (E- \epsilon - \lambda) -4 t^3 \cos ka \right] = 0
\label{disp2}
\end{equation}
\end{widetext}
From the above expression we can see that the network supports a dispersionless flat band mode at 
$E = \epsilon - \lambda$.

\section{Possible scope for experiment}
\label{expt}
In this section, it is very pertinent to highlight some scope for possible experimental realization
of flat photonic band in such quasi-one dimensional structures. The recent experiments by
S. Mukherjee \textit{et al.}~\cite{seba,seba2} and also by the other authors~\cite{vicen,hu,zong}
regarding the photonic analogue of localization of wave train in photonic lattices formed by laser-induced
single-mode wave guides prompt us to study the method of extracting the \textit{self-localized} modes
for different model networks. The present analysis discussed in this article 
can be experimentally verified
 in this era of pretty advanced nanotechnology and lithography
techniques. The wave propagation in such single-mode optical wave guides can be adjusted by
femtosecond laser-writing method as well as the optical induction technique. 
The overlap parameter can be manipulated by the dielectric properties of the core material.
This provides a scope for direct observation of diffraction free FB states.
Also, a synthetic gauge field~\cite{longhi1,longhi2}
 can be generated by modulating the propagation constant
to  study the effect of magnetic field in our proposed geometry.

\subsection{Photonic analogue}

Analogy between the electronic model and the corresponding photonic scenario within the
\begin{figure}[ht]
\centering
(a)\includegraphics[clip,width=6cm,angle=0]{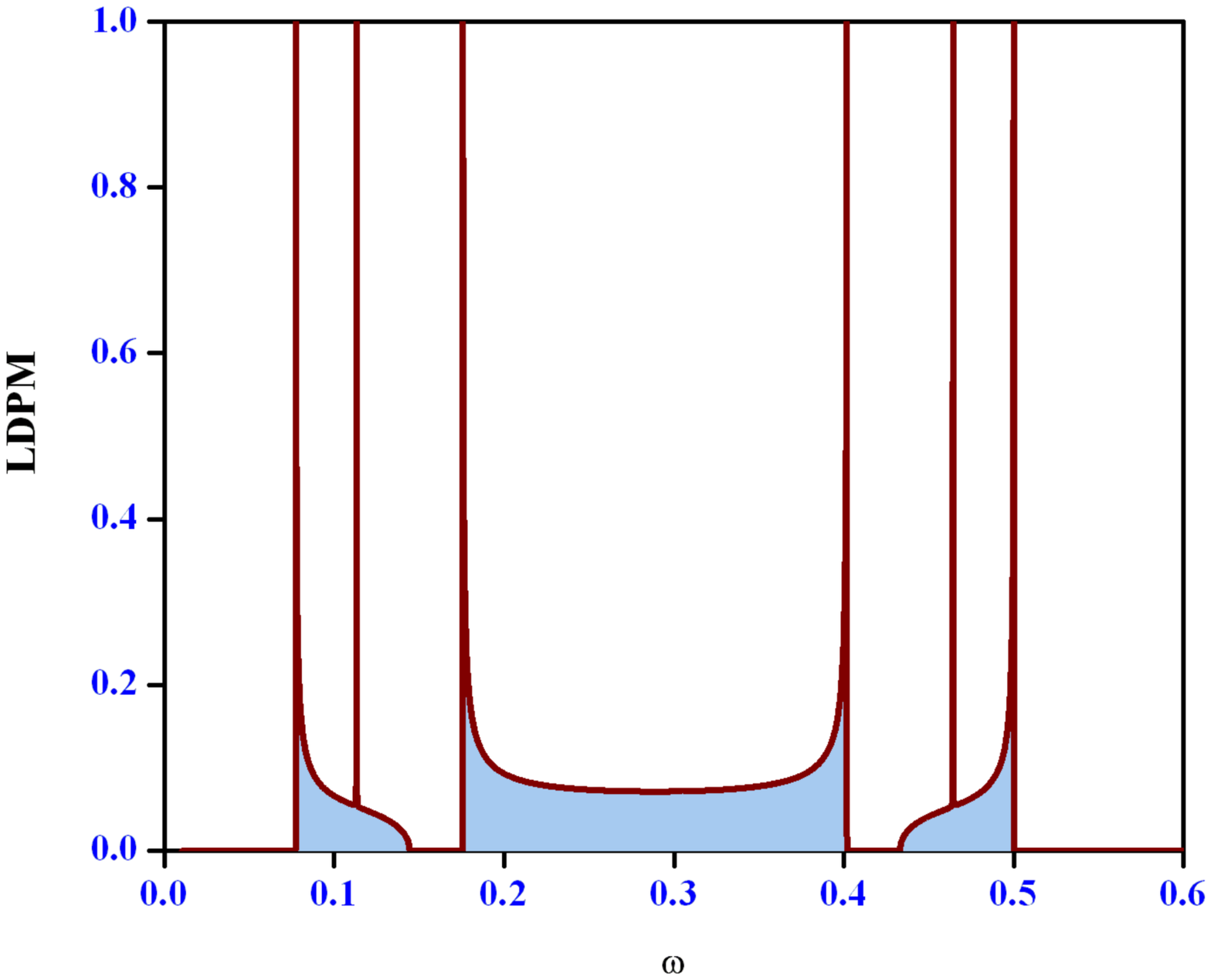}
(b)\includegraphics[clip,width=6cm,angle=0]{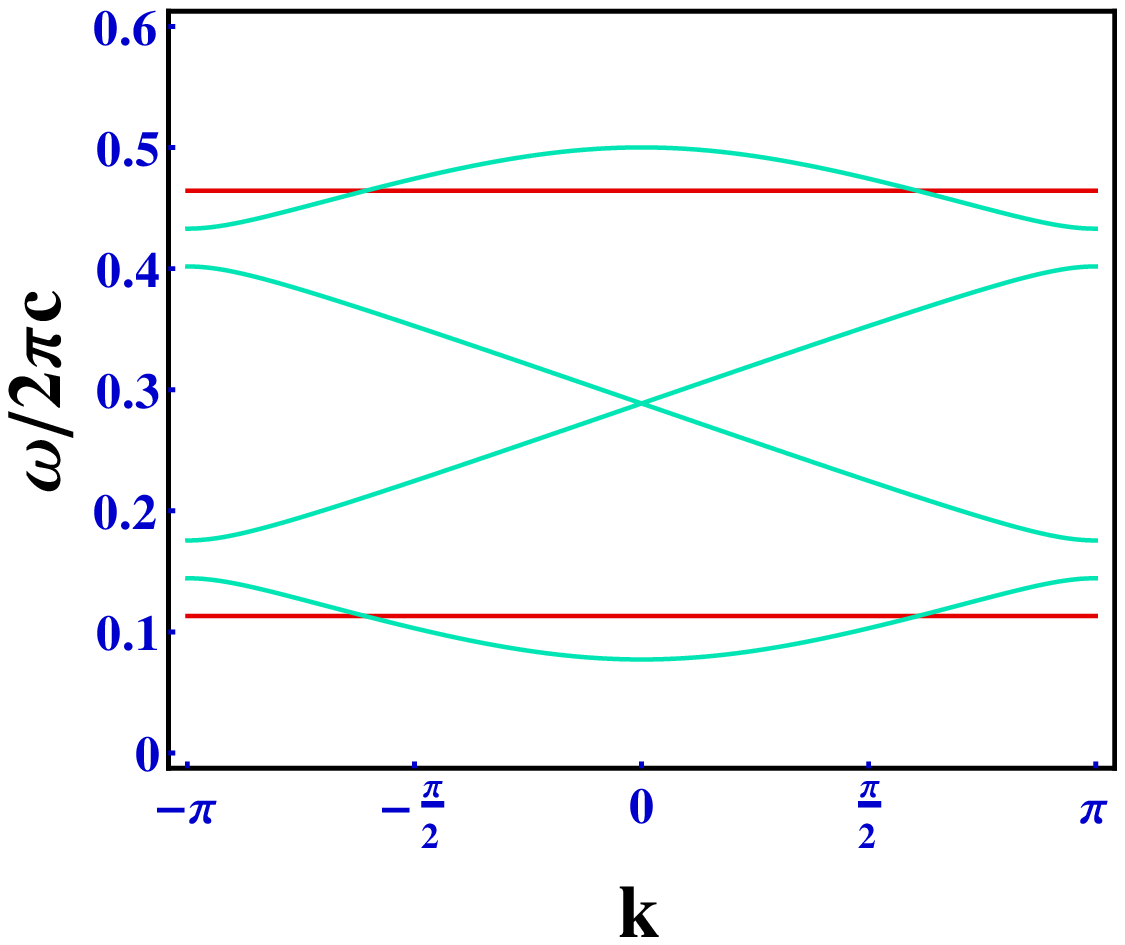}
\caption{(Color online) 
(a) Local density of photonic modes as a function of the frequency of the wave injected 
for the wave guide network arranged in $AB_2$ rhombic geometry and
(b) the corresponding optical dispersion relationship.}
\label{pyroopt}
\end{figure}
 tight-binding framework is not new. As first introduced by Sheng \textit{et al.}~\cite{wave1,wave2}, the wave propagation through any quasi-one dimensional network can be exactly mapped back onto the corresponding electronic case with the appropriate parametric substitution. 
The analogy~\cite{atanu4,biplabwg} of
 the network equations with that of an electron propagating in
 a similar lattice helps us to study the localization of classical waves. Also, this mapping is entirely a mathematical construction
  and once this is done the RSRG recursion relations are
 insensitive to whether the input comes from a quantum case
 or a classical one. Obviously here the propagation of classical wave can be manipulated by selective choice of dielectric parameter. If we set the length
 of the wave guide segment at the very outset, then one can
 easily compute the \textit{exact} frequency of injected wave train for which the self-localization occurs.

\subsection{Optical density of modes}
\label{DPM}
We have cited the distribution of allowed photonic modes using the tight-binding analogy.
\begin{figure}[ht]
\centering
(a)\includegraphics[clip,width=6cm,angle=0]{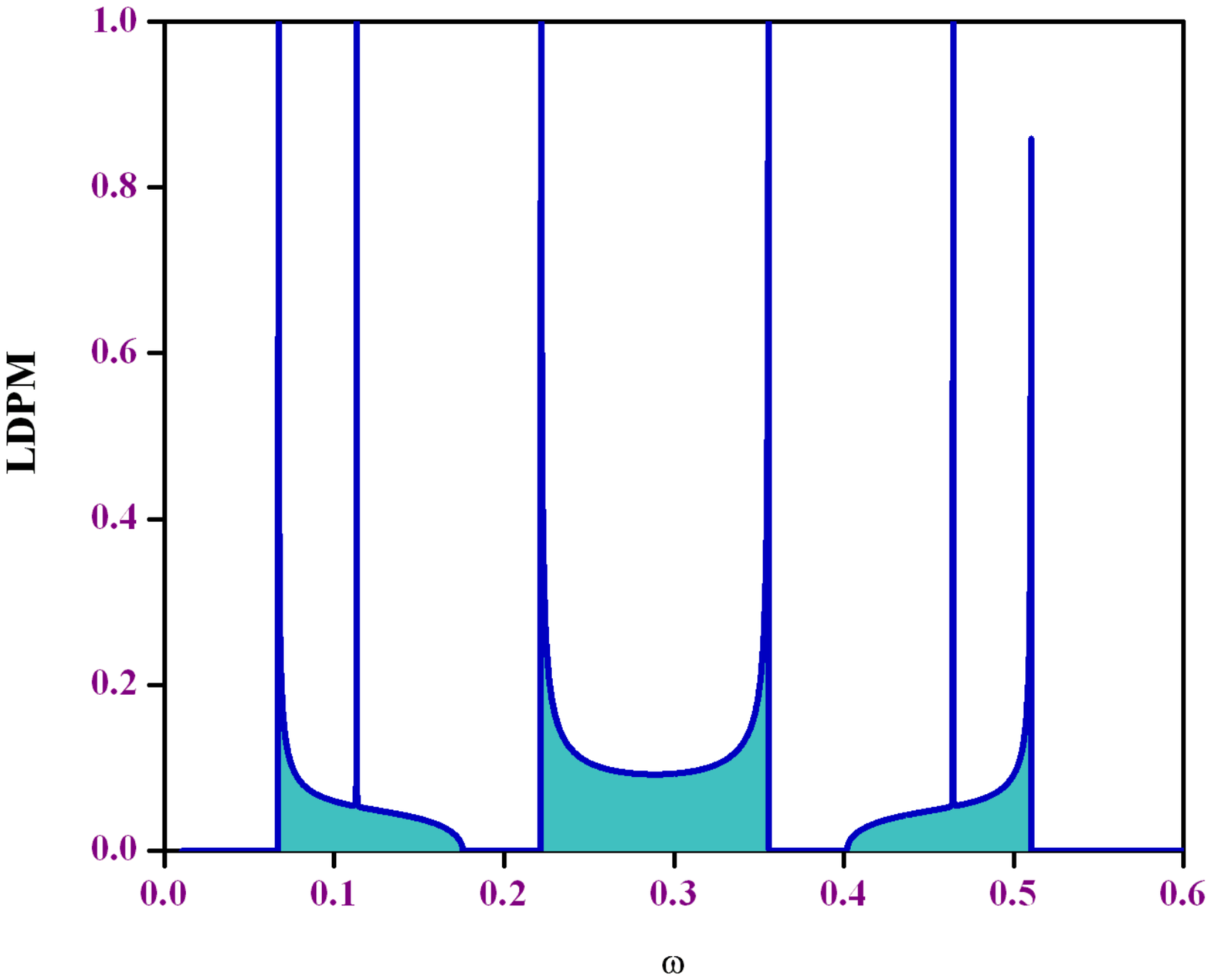}
(b)\includegraphics[clip,width=6cm,angle=0]{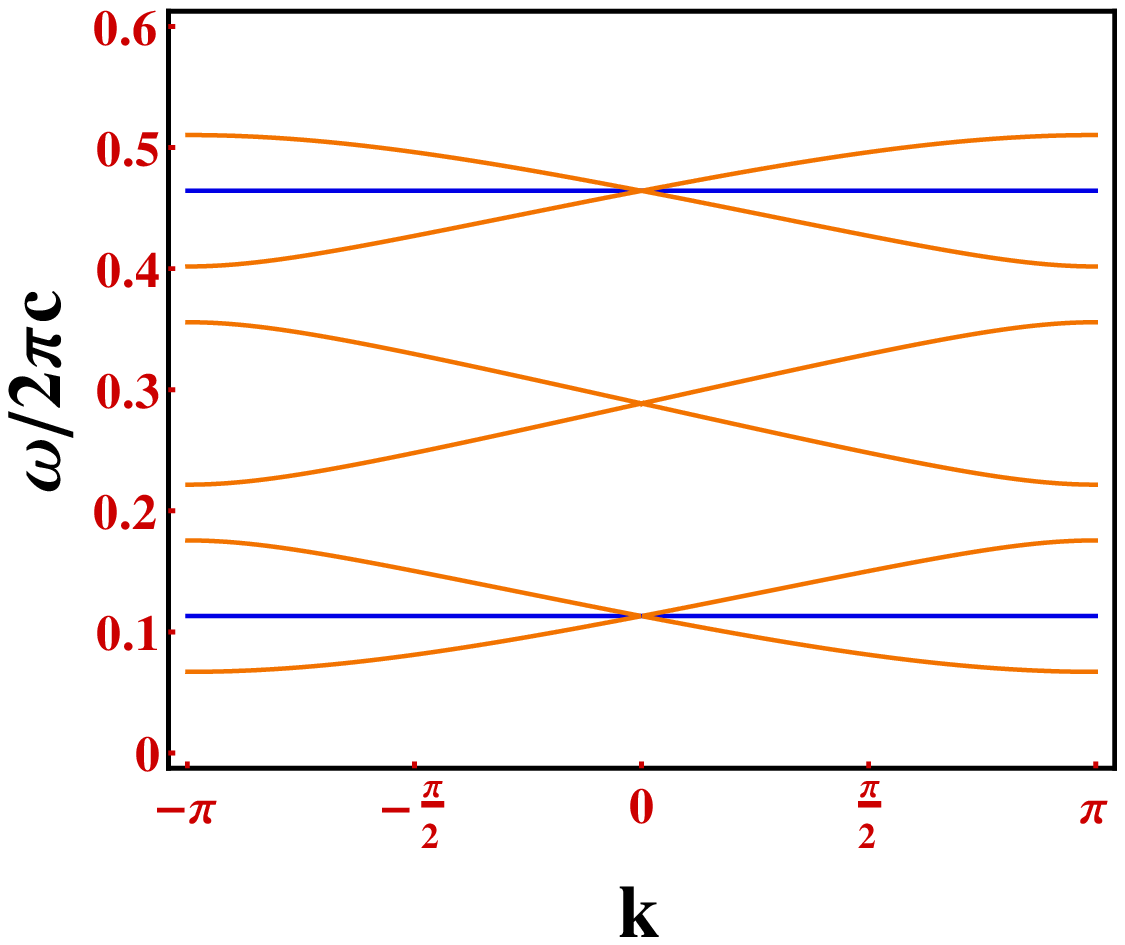}
\caption{(Color online) 
(a) Local density of photonic modes against the frequency of the signal 
for the wave guide network arranged in dimer-plaquette chain geometry and
(b) the corresponding photonic dispersion relationship.}
\label{dimeropt}
\end{figure}
The profile showing the local density of photonic modes (LDPM) is plotted
as a function of frequency of the injected signal within the range $0 < \omega/2 \pi c < 1/2$. The
plot shows clusters of nonzero values of LDPM over different subband regimes. It is needless to say
that quite arbitrarily we have taken the index of refraction of the core of the wave guide as
$n = \sqrt{3}$. Both the networks also exhibit gaps in between two subbands and thus can be a suitable 
candidate for photonic bandgap kind of system. Fig.~\ref{pyroopt}(a) and Fig.~\ref{dimeropt}(a) show the 
density of optical modes for rhombic lattice and dimer plaquette chain respectively.

\subsection{Flat photonic modes}

Because of zero group velocity, the corresponding single- 
particle state is sharply localized at a point, or in a finite cluster 
of nodal points in the system. Such clusters are separated from 
the neighboring clusters by vertices where the amplitude of the 
wave function is zero. Therefore, the movement of the particle 
is restricted along the periphery of those finite clusters.

By virtue of the decoupled equations and optical analogue of the 
tight-binding dispersion relation one can easily compute the flat 
photonic modes for the monomode wave guide network using the
one-to-one correspondence of electronic and photonic cases.
For $\epsilon_r = 3$, the non-dispersive roots are $\omega / 2 \pi c = 0.113, 0.465$.
Those two roots correspond to the left and right spikes appeared in
the DOS spectrum.

In photonics flat bands are closely related to the 
technologically-important concept of slow light, 
where the significant decrease in the wave group velocity offers 
enhanced nonlinear effects and is useful for pulse buffering.
If a single optical mode is non-dispersive in one direction, 
it implies localization in that direction as an extreme case,
unfolding the possibility of engineering ultraslow light. This 
is the ultimate consequence of a perfect geometric phase cancellation which 
causes effective mass tensor to be divergent, 
leading to a possible observation of ``heavy photons". This 
does not appear improbable if we borrow the language for the 
electrons, thanks to the analogous tight-binding model which 
indeed works quite accurately for the low-lying photonic 
dispersion. Vanishing curvature in the frequency ($\omega$)- vs - wave vector ($k$) 
curves signifies infinite effective mass of the particle involved. 
This in turn makes the mobility vanish. Due 
to the vanishing group velocity of the \textit{heavy photon}
the curvature of the band corresponding to those modes becomes completely flat in nature.

\subsection{Slow light and mode crossover}
\label{mode}
Light plays a very important as well as effective role as a messenger in the communication system and
optical communication is one of the well-studied aspect in the technological field. 
Slow light with a remarkably low group velocity of the wave train is a very promising solution for
optical delay line or optical buffering and advanced time-domain optical signal processing. It is also
anticipated to enhance linear and nonlinear effects~\cite{baba}
 and so miniaturize functional photonic devices, as slow
light compresses optical energy in space that eventually increases the light-matter interaction. This in
turn, results in enhanced gain and absorption, phase shift, and non-linearities. The flat band networks as
discussed in the previous section support one or more dispersionless bound states. For those specific states
the particle is locked in space. This observation gives oxygen to the photonic flat band researchers who
try to achieve a slow light condition by enforcing destructive interference. By virtue of this 
phase cancellation induced by topological symmetry present in the underlying structure,
the velocity of the light can be slowed down considerably~\cite{oka}. This amounts to a
photonic realization of the flat-band model in solid-state physics as conceived by Lieb~\cite{lieb},
 Mielke~\cite{mielke}
and Tasaki~\cite{tasaki}. 

The vanishing curvature of the flat band indicates extremely low group velocity.
The dense packing of the flat and the dispersive bands
therefore opens up a possibility of 
slow light engineering and re-entrant intermodal crossover of optical eigenmodes, going
from a resonant or diffusive (dispersive) state to a sharply localized or bound (flat) one,
or vice versa as one climbs the frequency axis at any fixed
value of the wave vector $k$ Such a crossover can, in principle,
be manipulated over arbitrarily small intervals of frequency.
 The entire network can thus
behave like a optical switch which selectively goes off (for a
localized, non-dispersive bound state) and on (for a resonant 
state) over arbitrary small frequency intervals.

\section{Concluding remarks}
\label{conclu}
We have shown an analytically exact scheme to extract the eigenvalues corresponding
 to the compact localized, non-dispersive, degenerate flat band eigenstates
 of a set of quasi-one dimensional structures. A real space renormalization group scheme is exploited
 to unravel such dispersionless sharply localized states. The evaluation of spectral landscape is corroborated by the numerical calculation of two-terminal transport. The flat curvature of energy band corresponding to those \textit{self-localized} is demonstrated using the exact computation of dispersion profile within the same tight-binding formalism. 

One-to-one mapping of electronic scenario onto the corresponding optical case prompts us to identify the analogous photonic flat band modes for the same geometry arranged in a single-mode optical wave guide structure. Such results bring an opportunity to modulate experimentally the localization of classical waves, for example, light,
 triggered by the lattice topology without bothering about the
 high permittivity of the core materials. It might be useful
 in developing novel photonic band-gap structures. Also, the
 physical significance of vanishing group velocity is that several
 scattering waves form standing wave pattern in such photonic
 wave guide network. For such self-localized eigenmodes, we
 can obtain coherent waves, i.e., lasing action~\cite{noda,timko} at the
 photonic band edges.
\begin{acknowledgments}
A.N. is thankful for the stimulating
discussions regarding the results with Prof. Arunava Chakrabarti.
\end{acknowledgments} 


\begin{thebibliography}{60}

\bibitem{anderson} P. W. Anderson, Phys. Rev. \textbf{109}, 1492 (1958).

\bibitem{suther} B. Sutherland, Phys. Rev. B \textbf{34}, 5208 (1986).

\bibitem{lieb} E. H. Lieb, Phys. Rev. Lett. \textbf{62}, 1201 (1989).

\bibitem{miel1} A. Mielke, J. Phys. A \textbf{24}, L73 (1991).

\bibitem{miel2} A. Mielke, J. Phys. A \textbf{24}, 3311 (1991).

\bibitem{tasaki} H. Tasaki, Phys. Rev. Lett. \textbf{69}, 1608 (1992).

\bibitem{miel3} A. Mielke and H. Tasaki, Commun. Math. Phys. \textbf{158}, 341 (1993).

\bibitem{tasaki2} H. Tasaki, Prog. Theor. Phys. \textbf{99}, 489 (1998).

\bibitem{kapit} E. Kapit and E. Mueller, Phys. Rev. Lett. \textbf{105}, 215303 (2010).

\bibitem{tang} E. Tang, J.-W. Mei, and X.-G. Wen, Phys. Rev. Lett. 106, 236802 (2011).

\bibitem{sun} K. Sun, Z. Gu, H. Katsura, and S. Das Sarma, Phys. Rev. Lett.
\textbf{106}, 236803 (2011).

\bibitem{chamon} T. Neupert, L. Santos, C. Chamon, and C. Mudry, Phys. Rev.
Lett. \textbf{106}, 236804 (2011).

\bibitem{yao} W. Yao, S.A. Yang, Q. Niu, Phys. Rev. Lett. \textbf{102}, 096801 (2009).

\bibitem{bloch} I. Bloch, J. Dalibard, W. Zwerger, Rev. Mod. Phys. \textbf{80}, 885 (2008).

\bibitem{lede} D.N. Christodoulides, F. Lederer, Y. Silberberg, Nature \textbf{424}, 817 (2003).

\bibitem{kim} N. Masumoto, N.Y. Kim, T. Byrnes, K. Kusudo, A. L\"{o}ffler, S. H\"{o}fling, A. Forchel,
Y. Yamamoto, New J. Phys. \textbf{14}, 065002 (2012).

\bibitem{dias} R. G. Dias and J. D. Gouveia, Sci. Rep. \textbf{5}, 16852 (2015).

\bibitem{mora} L. Morales-Inostroza and R. A. Vicencio, Phys. Rev. A \textbf{94}, 043831 (2016).

\bibitem{flach1} A. Ramachandran, A. Andreanov, and S. Flach, Phys. Rev. B \textbf{96}, 161104 (2017).

\bibitem{xu} C. Xu, G. Wang, Z. H. Hang, J. Luo, C. T. Chan, and Y. Lai, Sci. Rep. \textbf{5}, 18181 (2015).

\bibitem{flach2} S. Flach, D. Leykam, J. D. Bodyfelt, P. Matthies, and A. S. Desyatnikov, 
Europhys. Lett. \textbf{105}, 30001 (2014).

\bibitem{flach3} W. Maimaiti, A. Andreanov, H. C. Park, O. Gendelman, and
S. Flach, Phys. Rev. B \textbf{95}, 115135 (2017).

\bibitem{shen} S.-Q. Shen, Z.-M. Qiu, and  G.-S. Tian, Phys. Rev. Lett. \textbf{72}, 1280 (1994). 

\bibitem{sarma} C. Wu, D. Bergman, L. Balents, and S. D. Sarma, Phys. Rev. 
Lett. \textbf{99}, 070401 (2007). 

\bibitem{kawa} S. Miyahara, S. Kusuta, and N. Furukawa,  Physica C \textbf{460}, 1145 (2007). 

\bibitem{julku} A. Julku, S. Peotta, T. I. Vanhala, D.-H. Kim, and P. T\"{o}rm\"{a}, Phys. Rev. 
Lett. \textbf{117}, 045303 (2016). 

\bibitem{kopnin} N. B. Kopnin, T. T. Heikkil\"{a}, and G. E. Volovik, Phys. 
Rev. B \textbf{83}, 220503 (2011).

\bibitem{seba} S. Mukherjee, A. Spracklen, D. Choudhury, N. Goldman, P. \"{O}hberg,
E. Andersson, and R. R. Thomson, Phys. Rev. Lett.
\textbf{114}, 245504 (2015).

\bibitem{vicen} R. A. Vicencio, C. Cantillano, L. Morales-Inostroza, B. Real, 
C. Mej\'{i}a-Cort\'{e}s, S. Weimann, A. Szameit, and M. I. Molina, Phys.
Rev. Lett. \textbf{114}, 245503 (2015).

\bibitem{oka} S. Endo, T. Oka, and H. Aoki, Phys. Rev. B \textbf{81}, 113104 (2010).

\bibitem{atanu4} A. Nandy and A. Chakrabarti, Phys. Rev. A \textbf{93}, 013807 (2016).

\bibitem{goda} M. Goda, S. Nishino, and H. Matsuda, Phys. Rev. Lett. \textbf{96}, 126401 (2006).

\bibitem{miel4} A. Mielke, J. Phys. A-Math. Gen. \textbf{25}, 4335 (1992).

\bibitem{atanu2} A. Nandy, B. Pal, and A. Chakrabarti, J. Phys.: Condens. Matt. \textbf{27}, 125501 (2015).


\bibitem{mejia} R. A. Vicencio and C. Mej\'{i}a-Cort\'{e}s, J. Opt. \textbf{16}, 015706 (2014). 

\bibitem{roja} S. Rojas-Rojas, L. Morales-Inostroza, R. A. Vicencio, and 
A. Delgado, arXiv:1706.01988 (2017). 

\bibitem{xia} S. Xia, Y. Hu, D. Song, Y. Zong, L. Tang, and Z. Chen, Opt. 
Lett. \textbf{41}, 1435 (2016).

\bibitem{mejia2} R. A. Vicencio, C. Cantillano, L. Morales-Inostroza, B. Real, 
C. Mej\'{i}a-Cort\'{e}s, S. Weimann, A. Szameit, and M. I. Molina, 
Phys. Rev. Lett. \textbf{114}, 245503 (2015). 

\bibitem{von} J. von Neuman and E. Wigner, Phys. Z. \textbf{30}, 467 (1929). 

\bibitem{herr} F. H. Stillinger and D. R. Herrick, Phys. Rev. A \textbf{11}, 446 (1975).

\bibitem{ringel} Y. E. Kraus, Y. Lahini, Z. Ringel, M. Verbin, and O. Zilberberg,
Phys. Rev. Lett. \textbf{109}, 106402 (2012).

\bibitem{sorel} Y. Lahini, R. Pugatch, F. Pozzi, M. Sorel, R. Morandotti, N.
Davidson, and Y. Silberberg, Phys. Rev. Lett. \textbf{103}, 013901
(2009).

\bibitem{ochi} M. Ochi, R. Arita, M. Matsumoto, H. Kino, and T. Miyake, Phys. Rev. B \textbf{91}, 165137 (2015). 

\bibitem{hase} I. Hase, T. Yanagisawa, and K. Kawashima, Nanoscale Res. Lett. \textbf{13}, 63 (2018).

\bibitem{flach4} J. D. Bodyfelt, D. Leykam, C. Danieli, X. Yu, and S. Flach, Phys.
Rev. Lett. \textbf{113}, 236403 (2014).

\bibitem{mati} M. Hyrk\"{a}s, V. Apaja, and M. Manninen, Phys. Rev. A \textbf{87}, 023614
(2013).

\bibitem{greenfunc} S. Souma and A. Suzuki, Phys. Rev. B \textbf{65}, 115307 (2002).

\bibitem{liu} Y. Liu and K. A. Chao, Phys. Rev. B \textbf{34}, 5247 (1986).

\bibitem{derz} O. Derzhko, J. Richter, O. Krupnitska, and T. Krokhmalskii, Phys. Rev. B \textbf{88}, 094426 (2013).

\bibitem{seba2} S. Mukherjee and R. R. Thomson, Opt. Lett. \textbf{40}, 5443
(2015).

\bibitem{hu} S. Xia, Y. Hu., D. Song, Y. Zong, L. Tang, and Z. Chen,
Opt. Lett. \textbf{41}, 1435 (2016).

\bibitem{zong} Y. Zong, S. Xia, L. Tang, D. Song, Y. Hu, Y. Pei, J. Su,
Y. Li, and Z. Chen, Opt. Express \textbf{24}, 8877 (2016).

\bibitem{longhi1} S. Longhi, Opt. Lett. \textbf{39}, 5892 (2014).

\bibitem{longhi2} S. Longhi, Opt. Lett. \textbf{38}, 3570 (2013).

\bibitem{wave1} S. Alexander, Phys. Rev. B \textbf{27}, 1541 (1983).

\bibitem{wave2} Z.-Q. Zhang and P. Sheng, Phys. Rev. B \textbf{49}, 83 (1994).

\bibitem{biplabwg} B. Pal, P. Patra, J. P. Saha, and A. Chakrabarti, Phys. Rev. A \textbf{87}, 023814 (2013).

\bibitem{flachreview} D. Leykam, A. Andreanov, and S. Flach, Adv. Phys.: X \textbf{3}, 1473052 (2018).

\bibitem{baba} T. Baba, Nat. Photonics \textbf{2}, 465 (2008).

\bibitem{oka} S. Endo, T. Oka, and H. Aoki, Phys. Rev. B \textbf{81}, 113104 (2010).

\bibitem{mielke} A. Mielke, Phys. Lett. A \textbf{174}, 443 (1993).

\bibitem{noda} M. Imada, S. Noda, A. Chutinan, T. Tokuda, M. Murata, and G. Sasaki, 
Appl. Phys. Lett. \textbf{75}, 316 (1999).

\bibitem{timko} M. Meier, A. Mekis, A. Dodabalapur, A. Timko, R. E. Slusher,
J. D. Joannopoulos, and O. Nalamasu, Appl. Phys. Lett. \textbf{74}, 7 (1999).













\end{thebibliography}
\end{document}